# DOES THE CORONA BOREALIS SUPERCLUSTER FORM A GIANT BINARY-LIKE SYSTEM?

**Giovanni C. Baiesi Pillastrini**[1*]

**ABSTRACT**

The distribution of local gravitational potentials generated by a complete volume-limited sample of galaxy groups and clusters filling the Corona Borealis region has been derived to search for new gravitational hints in the context of clustering analysis unrevealed by alternative methodologies. Mapping such a distribution as a function of spatial positions, the deepest potential wells in the sample trace unambiguously the locations of the densest galaxy cluster clumps providing the physical keys to bring out gravitational features connected to the formation, composition and evolution of the major clustered structures filling that region. As expected, the three deepest potential wells found at Equatorial coordinates: (~ 230°, ~ 28°, $z$ ~ 0.075), (~ 240°, ~ 27°, $z$ ~ 0.09) and, (227°, 5.8°, $z$ ~ 0.0788) correspond to massive superclusters of galaxy groups and clusters identified as the Corona Borealis, A2142 and Virgo-Serpent, respectively. However, the deepest isopotential contours around the Corona Borealis and A2142 superclusters seem to suggest a gravitational feature similar to a giant binary-like system connected by a filamentary structure. To a first approximation, it seems unlikely that this hypothesized system could be gravitationally bound.

[1] *U.A.I. c/o Osservatorio Astronomico Fuligni - Via Lazio 14, 00040 Rocca di Papa (RM), Italy*
[*] *permanent address: via Pizzardi, 13 - 40138 Bologna - Italy - email: gcbp@it.packardbell.org*

# 1. INTRODUCTION

## 1.1. The Corona Borealis region (CBr)

Studying the distribution and dynamics of galaxy superclusters in the local Universe, Bahcall and Soneira (1984) and, more recently, Luparello et al. (2011) analyzing the Corona Borealis region (CBr hereafter) hypothesized that the well-known Corona Borealis Supercluster (CBSCL hereafter) is part of a much more extended and massive structure. Stimulated in disentangling this issue, we attempt an exploratory analysis of that region based on the gravitational potential method (GPM hereafter; Baiesi Pillastrini 2013) with the main aim to search new gravitational hints and features unrevealed by previous studies as well as to compare the efficiency of the GPM in identifying and quantifying clustered structured with the results of previous well-known studies. Since the first identification of the CBSCL by Abell (1961) using his own Catalog of Galaxy Clusters (Abell 1958), that region has been largely investigated using a variety of clustering algorithms generally based on the density field and Friend of Friend (FoF) analyses (Bahcall and Soneira 1984; Cappi and Maurogordato 1992; Zucca et al. 1993; Kalinkov and Kuneva 1995 Einasto et al. 1994, 1997, 2001, 2011a) and compared with the Abell cluster Catalog. On the other hand, many other dedicated studies have analyzed its composition, morphology and dynamical state (Postman et al. 1988; Small et al. 1997, 1998; Kopylova and Kopylov 1998; Marini et al. 2004; Génova-Santos et al. 2010; Batiste and Batuski 2013; Pearson et al. 2014; Einasto et al. 2015; Gramann et al. 2015; Pearson 2015). A new generation of Supercluster catalogs constructed with accurate and complete datasets combined with new methodologies of the clustering analysis has provided insight on the extension and membership of the CBSCL ( Einasto et al. 2006; Luparello et al. 2011; Liivamagi et al. 2012; Chow-Martinez et al. 2014).

## 1.2. Clustering algorithms vs. GPM

The common practice of introducing selection parameters depending on well-motivated assumptions in the clustering algorithms and analyses such as linking lengths, spatial density thresholds, etc., often provides quite different boundary and membership to a certain structure. For example, the Abell clusters assignment to the CBSCL was subject to many revisions after the first definition of Abell (1961). In the present study, the GPM clustering algorithm based on the Newtonian gravity theory has been applied in order to detect the major clustered structures in the Corona Borealis region, their main physical properties and, if any, unknown gravitational features. The GPM was developed following the prescription of the exploratory data analysis and rests on the basic idea that the gravitational potential is closely connected with the matter density field and that galaxy systems aggregate by following the laws of gravity no matter how different they are. As established by the theory of gravitational instability, the formation (and evolution) of huge scale structures seen in the galaxy distribution is tightly related to the potential field distribution (Madsen et al.1998). It follows that clustered regions arise due to slow matter flows into negative potential wells so that, the detection of huge mass concentrations can be carried out simply observing the regions where the *deepest potential wells* (DPW hereafter) originate. Its application is becoming now possible after that accurate mass estimations become available in large galaxy group/cluster catalogs up to intermediate redshift (see for instance Tempel et al. 2014). The use of large datasets of galaxy systems taken as *mass tracers of gravitational potential wells* is the most relevant difference between the GPM and alternative methods based on the analysis of space density or velocity fields. The GPM was designed to construct analytically a list of the deepest potential magnitudes of a complete volume-limited dataset of astronomical objects and, graphically, to display isopotential contours from which one can explore and identify the location of a single or more clustered structures simply looking for the deepest negative potential counterparts. Specifically, the GPM performs a two-step analysis as follows: after the identification in position and in magnitude of the DPWs, each DPW is assumed as the temporary center of mass then, by modeling an appropriate mass-radius relation, the quantitative parameters defining the mass overdensity can be iteratively computed until the final position of the center of mass remain constant.

In the present study we assume: $H_0 = 100\ h\ km\ s^{-1}\ Mpc^{-1}$, $\Omega_m = .27$ and $\Omega_\Lambda = .73$ according to the cosmological parameters of the dataset used hereafter.

The paper is organized as follows: in Sect.2 we briefly describe the GPM. In Sect.3 the GPM is applied to a complete volume-limited sample of galaxy groups and clusters filling the CBr with the purpose to identify the locations of the DPWs. Then, in Sect.4, the assumed criterion to quantify the mass distribution underlying the DPW clumps is described and applied. In Sect.5 the results are then compared with other studies. In Sect.6, the gravitational binding of the proposed binary system is tested. In Sect.7, conclusions are drawn.

# 2. A BRIEF DESCRIPTION OF THE GRAVITATIONAL POTENTIAL METHOD (GPM)

## 2.1. The algorithm design in the framework of the ΛCDM cosmological model

The methodology of investigation adopted for the GPM is essentially based on the exploratory data analysis (Tukey 1977) in the framework of Newtonian mechanics with the aim to construct the *local* gravitational potential distribution generated by a complete volume-limited sample of astronomical objects. Now, being gravity a *superposable force,* the gravitational potential generated by a collection of point masses at a certain location in space is the sum of the potentials generated at that location by each point mass taken in isolation. By measuring the local potential at the position of each object taken one at a time as a test-particle, the map of the local potential distribution generated by the spatial distribution of the whole sample is displayed. *The DPWs identify unambiguously the location of the densest clumps in a mass distribution.* Now, in the framework of the ΛCDM cosmological model, the total potential acting on a test-particle is given by $\Phi = -U_g - U_\Lambda$ where $U_g$ is the *attractive component* of the potential due to gravity and $U_\Lambda$ is the *repulsive component* of the potential due to dark energy. Given $N_{V_j}$ point-masses located at position vectors $d_i$ (from the observer) within a spherical volume $V_j$ of fixed radius $R_V$ centered on a generic test-particle $j$ at position vector $d_j$ from the observer then, potential components generated at position vector $d_j$ by the $N_{V_j}$ point masses $m_i$ ($i = 1,...., N_{V_j}$) are given by

$$U_g = G \sum_{i=1, i \neq j, i \in V_j}^{N_{V_j}} m_i (d_i - d_j)^{-1} \quad \text{and} \quad U_\Lambda = \frac{4\pi}{3} \rho_\Lambda G \sum_{i=1, i \neq j, i \in V_j}^{N_{V_j}} (d_i - d_j)^2$$

where $G$ is the gravitational constant and $\rho_\Lambda$ is dark energy density of ~6 x 10$^{-30}$ g/cm$^3$ (Plank collaboration, 2015). Repeating the calculation for each point-mass taken one at a time as a test-particle, we construct the whole $\Phi_j$ distribution. $\Phi_j$ are given in $10^6 h(km/s)^2$ unit and is always $\leq 0$.

Since the GPM is a *gravity*-based method to detect gravitational clustering, for each point-mass $j$ where the inequality $|U_\Lambda| > |U_g|$ is satisfied, $\Phi_j$ is assumed = 0. This assumption is required to prevent objects dominated by the repulsive potential component to mixed up opposite actions as gravitational attraction and dark energy repulsion. An object dominated by the repulsive potential component must follow the accelerated expansion of the Universe, so that it cannot be taken into account in the clustering analysis to define bound structures. These objects will represent the *zero-level* of the $\Phi_j$ distribution so that only features subject to gravitational attraction will be highlighted removing fake images and enhancing high-resolved images of real clustered structures.

**2.2. Advantages and disadvantages**

The GPM provides several relevant advantage: *i)* it enables the identification of clustered structures using an algorithm based on gravity theory; *ii)* being gravity a long range force, the potential distribution is smoother than the density distribution since the contribution to local potential fields due to small density fluctuations is irrelevant *e.g.* galaxy pairs and triplets; *iii)* gravity-based selection algorithm enables to constrain overdensities with a clearer physical meaning than, for example, spatial density-based algorithms that are independent from gravitational influences and interactions. The main disadvantage of the GPM is that its accuracy in detecting superstructures depends largely on the accuracy of mass estimations. In other words, the more accurate are the assumed mass estimates provided by a certain dataset, the more reliable the clustering analysis will be. It follows that the GPM applied to different datasets constructed with different mass estimates, spatial reconstruction techniques or different selection methods, may give different results.

**3. THE GPM APPLIED TO THE CORONA BOREALIS REGION**

**3.1. The dataset**

Each clustering algorithm can be accurate if the selected sample of objects under study is a complete volume-limited and free of bias effects (selection effect, redshift distortion and so on). In studies concerning gravitational interactions, the use of cluster samples overcomes some of these problems faced, for example, by galaxy samples since clusters are luminous enough for samples to be volume-limited out to large distances, trace the peaks of the density fluctuation and reduce the effect of redshift distortion. Therefore, a galaxy cluster sample emerges as the most convenient mass tracer candidate for the present clustering analysis. In particular, the best choice would be a complete volume-limited catalog of galaxy clusters where reliable mass estimations (assumed as point-mass tracers) are available.

Recently, Tempel et al. (2014, T14), applied an improved Friends of Friends (FoF) method to flux- and volume-limited galaxy samples drawn from the SDSS DR10 survey (Ahn et al. 2014) main contiguous area covering 7221 square degrees in the sky. It has been used to trace groups and clusters of galaxies out to $z = 0.2$ involving 588,193 galaxies with spectroscopic redshift. Their technique provided a flux-limited catalog of over 82,458 galaxy group/clusters and, seven other catalogs constructed volume-limited with different absolute magnitude limits: from M = -18 to -21. The M = -20 volume-limited catalog has been adopted here. It lists 24,258 galaxy group/clusters which has been used in the present analysis. For each identified cluster, the catalog list the following parameters of our interest: ID of each object, n° of galaxy of the group/cluster, J2000 equatorial coordinates of the center as the origin, spectroscopic redshift (CMB-corrected), comoving distance in $h^{-1}Mpc$ and, finally, the estimated dynamical mass (assuming NFW profile) in solar mass unit. From this sample, a subsample of 6373 group/clusters filling the Corona Borealis region constrained by: 200° < R.A. < 260°, 0°< Decl. < 40° and, radially, from the comoving distance of 163 $h^{-1}Mpc$ (z = 0.055) to the limit for completeness of 322.6 $h^{-1}Mpc$ (T14) has been selected. T14 warn of the large error affecting the mass estimation of the galaxy pairs and triplets therefore, to reduce the bias due to outliers, all pairs have been removed from the subsample retaining 2,809 systems with n° ≥ 3 galaxies. Triplets have been retained to guarantee a high-resolved display of the isopotential contour levels with the condition of using the median mass of the sample as a proxy for their mass estimates.

### 3.2. Simplifying assumptions

*i)* The GPM assumes that the gravitational potential is *time-independent*;

*ii)* To overcome the problem of finding a finite solution of $\Phi_j$ for infinite gravitating masses, we need to assume the form of the spatial distribution of these masses. By considering that at the position of each test cluster, the local gravitational potential is mainly influenced by close neighbors and much less by distant masses *i.e.* $\Phi_j \to 0$ when $d_i - d_j \to \infty$, we may assume that the mass distribution within the spherical volume $V_j$ of fixed radius $R_V$ is embedded in a *uniform background*. Such supposed segregation of galaxy groups and clusters within $V_j$ provides the *finiteness* of the local gravitational potential. Outside $V_j$ the potential vanishes that is, at the distance of $d_i - d_j \geq R_v$, $\Phi_j \to 0$. For our purpose, to a first approximation, $V_j$ should be large enough to enclose the largest cluster concentration of the region in order to include their potential influence on the test-particles placed in its center. These massive objects generally fill volumes of about ~50 $h^{-1}Mpc$ radius then, $V_j$ should have a minimum radius larger than that, at least. Besides, this radius in addition to being large enough to prevent the so-called shot noise error and incorporates the major share of the gravitational influence exerted by neighboring masses, it should be large enough to avoid that the evaluation of $\Phi_j$ varies with $V_j$ of increasing $R_V$ more than the standard error of 16% on its amplitude (see Sect. 3.3). To verify this condition, one hundred random tests for increasing $R_V$ have been performed finding that at $R_V \sim 80$ $h^{-1}Mpc$ the variation of $\Phi_j$ is less than 14%.

*iii)* Triplets have been retained within the selected subsample of groups and clusters even if their mass estimates are affected by large uncertainty. This choice is justified for two reasons: first, removing them from the subsample we would have lost almost half of information about the local potential distribution and, second, they have very little influence on the potential determination: for example, a triplet with very large mass, say, of 5 x $10^{13}$ $M_\odot$ placed at an extreme short distance of 0.5 *Mpc* from a test-particle, adds to $\Phi_j$ a potential less than ~ 8% of a DPW. Furthermore, the median mass evaluated for the whole triplet sample has been assumed as a proxy in order to reduce the bias due to outliers (Einasto et al. 2015).

### 3.3. Uncertainties

The volume-limited group/cluster catalog of T14 does not provide errors associated to mass estimates. Fortunately, two recent studies of Old et al. (2014, 2015) analyzed errors in mass estimates comparing different mass estimation methods using simulated mock galaxy catalogues. According to their results, mass estimates listed in the T14 Catalog show ~ 50% scatter compared with their true values. By knowing that errors on spectroscopic redshifts of the SDSS DR10 survey do not exceed a few % (error due to cluster peculiar velocities is not take into account since smaller than that of a single galaxy) it is now possible evaluate statistically the uncertainty on $\Phi$. To quantify it, a Monte-Carlo simulation based on the resampling technique has been applied (Andrae 2010) to a random subsample enclosed in a spherical volume of $R_V = 80$ $h^{-1}Mpc$. Then, assuming a Gaussian error distribution of ~ 3% for spectroscopic redshifts and ~ 50% for cluster mass estimates, we can now randomly sample new data points to estimate the simulated $\Phi_j$ at the volume center. Repeating this resampling task 10,000 times, we get the distribution of the simulated data from which we can then infer the uncertainty given by the standard deviation. An estimated standard error of ~16% has been found which ensures a fair reconstruction of the local gravitational potential distribution.

## 3.4. Displaying the $\Phi_j$ distribution in 2D density contour plot

The outputs of the GPM routine consist of a numerical file where each cluster is identified by its 3D comoving position associated to the calculated $\Phi_j$ sorted by negative increasing values and a 2D contour map which displays the $\Phi_j$ distribution as a function of spatial positions integrated along the line of sight. This graphic tool enables to model certain qualitative aspects of the underlying mass distribution through appropriate choices of the number of isopotential contour levels. Fig.1 shows the $\Phi_j$ distribution highlighted by 6 contour levels filled with different colors from the zero-level ($\Phi_j = 0$, the red sea where below stay objects subject to gravitational attraction and above those following the local Hubble flow) to the DPW level ($\Phi_j \leq -1.768 \times 10^6 \; h(km/s)^2$, the dark blue peaks). As expected, Fig.1 shows three large and deep potential wells corresponding to well-known superclusters labeled in the Figure as CBSCL (Corona Borealis supercluster), A2142SCL (A2142 supercluster) and VIR-SER SCL (Virgo-Serpent supercluster) and their major Abell cluster members. At a first glance one can see a very interesting gravitational feature in the central part of the CBr where the two massive superclusters, the CBSCL and A2142SCL are dominant. They are spatially separated by a short distance less than 57 $h^{-1}Mpc$ (at their center of mass) but connected by a filamentary structure which seems to suggest a configuration similar to a binary system. Even if such a configuration is expected from the $\Lambda$CDM cosmological model which predicts that massive galaxy concentrations live at the intersection of large-scale filamentary structures generated through the merging of substructures lying along them (Plionis 2004), if confirmed, it would be the first observed case of a giant binary supercluster. It is worth noting that the CBSCL, A2142SCL and VIR-SER SCL lie almost in the same plane. This peculiar configuration was already identified by Einasto et al. (1997, 2011b) as part of a more extended plane named "supercluster plane" (in their papers, the three superclusters were named as the SCl 094, SCl 001 and SCl 011, respectively).

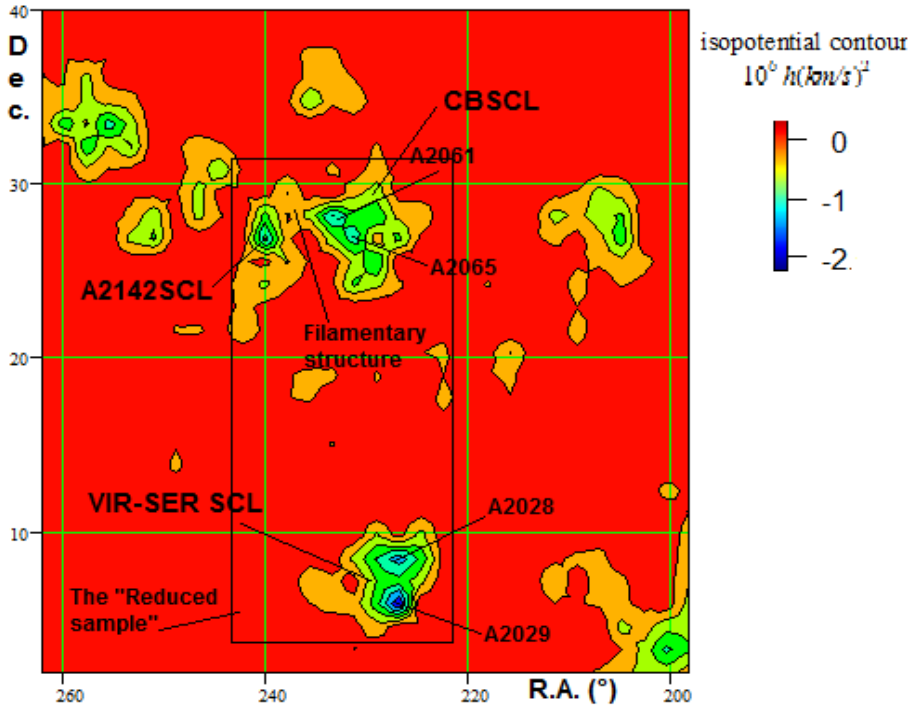

**Fig.1** Contour plot of the $\Phi_j$ projected density surface as a function of R.A. - Dec (in degree) plot. As expected, the deepest contours of the $\Phi_j$ distribution recover the three major superclusters of the Corona Borealis region: The CBSCL (Corona Borealis supercluster), A2142SCL and VIR-SER SCL (Virgo-Serpent supercluster). A filament connecting the CBSCL to the A2142SCL is apparent. For the Reduced sample area outlined in the Figure, see text.

## 3.5. Displaying the $\Phi_j$ distribution in 3D density isosurface plot

To be sure that isopotential density contours traced in Fig. 1 are not due to a graphical artifact, a detailed investigation of the mass distribution within the "binary" region is required. In other words, the clustering analysis should be restricted using a reduced subsample of objects extracted from the original dataset located within the binary region limited by $223° <$ R.A.$< 245°$, $4° <$ Dec $< 33°$ and $.069 < z < .0936$ or, in comoving distance, $200 < d < 280$ $h^{-1}Mpc$ (see Fig. 1). It is composed of 415 objects having $\Phi_j < 0$. In order to emphasize the gravitational features of the structures, from this sample a smaller one named the "Reduced sample" has been selected. It is composed of 217 objects having $\Phi_j \leq$ -0.4 x $10^6$ $h(km/s)^2$. From the gravitational point of view, the cutaway allows the selection of objects subject to the major gravitational influence induced by the environment then, "skeletons" drawn by the Reduced sample should represent the true bound structures of the region. In Fig. 2, the $\Phi_j$ distribution of that Reduced sample is displayed in a 3D Cartesian frame by dots drawn with a color scale varying with their negative magnitudes. The coordinate conversion from Equatorial (R.A., Dec) to Cartesian (x, y, z) has been obtained from $x = d \cdot cos(R.A.) \cdot cos(Dec.)$, $y = -d \cdot sin(R.A.) \cdot cos(Dec.)$ and $z = d \cdot sin(Dec.)$.

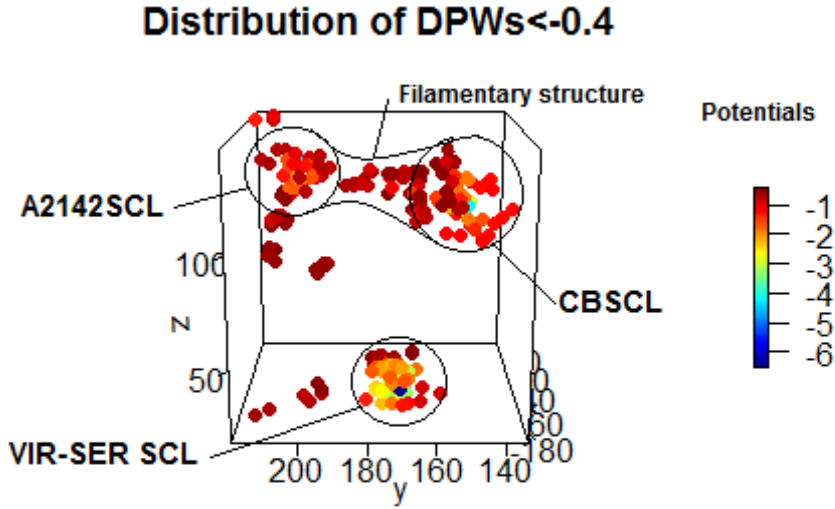

**Fig.2** The $\Phi_j$ (3D) distribution of the Reduced sample in the *x, y, z* frame is apparent. The deepest potential wells (dark and light blue colors) recover the position of the CBSCL, A2142SCL and VIR-SER SCL. A filamentary structure connects the CBSCL with the A2142SCL. Comoving distances are in $h^{-1}Mpc$ unit.

The binary-like system and the filamentary structure seen in Fig. 1 appear now well defined in a 3D visualization. Furthermore, using a 3D kernel density function, gravitational features emerging from Fig. 1 and 2 are now visualized in Fig. 3 as geometrical structures shaped by isosurface density contours of volume data . Again, even if the three superclusters turn out well defined by the isosurface density contours of the Reduced sample, from Fig.2 the filamentary structure departing from the CBSCL toward the direction of the A2142SCL seems incomplete suggesting that the binary members are not really connected. The spatial density estimates of the Reduced sample have been provided by a 3D kernel density function (kde3D) which uses a R-code (Feng and Tierney 2008). The function returns a three-dimensional array of estimated density values obtained from 40 grid points and a bandwidth of 3.5 $h^{-1}Mpc$ from which a 3D contour function displays the density isosurface contours at a certain level.

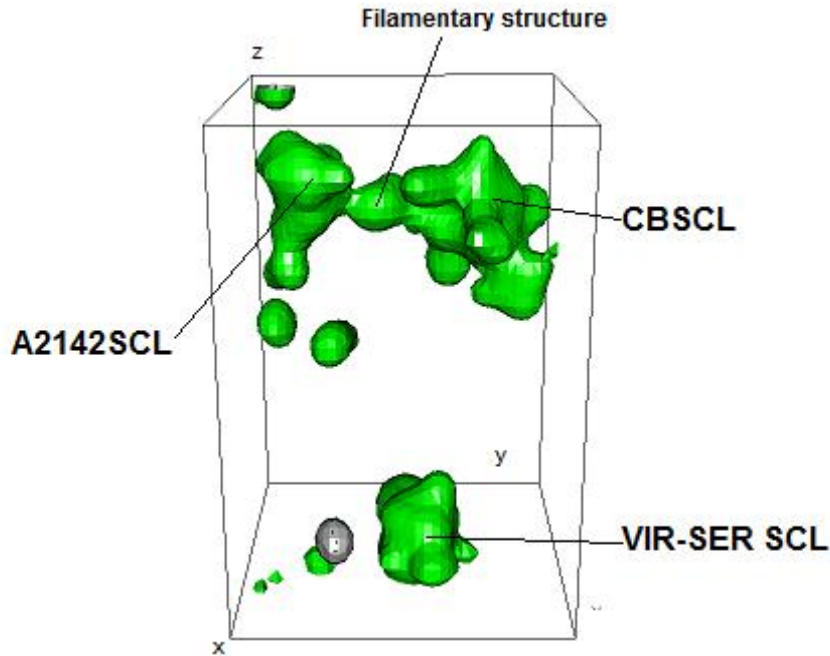

**Fig. 3** shows 3D density isosurface contours obtained from data points of the Reduced sample (see text). Isosurface contours overlap the scatter plot of data points filling the selected region. The superclusters CBSCL, A2142SCL and VIR-SER SCL as well as the filamentary structure connecting the CBSCL with the A2142SCL are well defined.

**4. SUPERCLUSTER EXTENTS, MASSES AND MEMBERSHIPS**

**4.1. On the definition of bound, collapsed, virialized structures**

The present analysis is devoted to identify large scale structures as well as unknown gravitational features among them. This is not an easy task since the definition of superclusters or superstructures is still not well established. As argued by Chon et al. (2015), these structures are generally defined as agglomerates of galaxy groups and clusters above a certain spatial density threshold without a clear definition and heterogeneous properties. Observationally, they are transition objects that largely reflect their initial conditions but unlike clusters, these structures are not virialized and have not reached a dynamical equilibrium. Therefore, a solution to correctly classify these objects is to include their future evolution selecting only those structures that will collapse in the future in a more homogeneous class of objects. The future evolution of a structure can be predicted using a model able to describes with reasonable accuracy every phase of its evolution as well as its dynamical state (Gramann and Suhhonenko 2002, Dünner et al. 2006, Luparello et al. 2011). Recently, many relevant studies improved the definition of superclusters (Chon et al. 2015; Teerikorpi et al.2015; Einasto et al.2015; Gramann et al. 2015; Pearson 2015). Almost all of these studies are based on the spherical collapse model which, in many case, could be a good approximation but not in general as pointed out by Einasto et al. (2015) warning to beware of the use of it to define the dynamical state of *anisotropic* structures as, for example, the A2142 supercluster without a detailed analysis of the internal density distribution. In conclusion, to achieve a precise and detailed knowledge of the physical properties of large scale structures one should have achieved many observational parameters, a task which is beyond the aim of the present study. Therefore, we will limit our effort applying a model which, at large, can give the essential structural parameters to define the superclusters under study.

**4.2. The adopted model**

One aim of this study is to compare the GPM efficiency in finding clustered structures with counterparts defined by more accurate studies. Then, we need a model to quantify mass and extent of each supercluster without claiming to be rigorous and exhaustive in defining their dynamical states. The simple way to constrain the mass and extent of a structure can be provided by the *maximum turnaround-mass relation* predicted in the framework of the ΛCDM cosmological model (Dunner et al. 2006, Chernin et al. 2009, Merafina et al. 2014, Pavlidou and Tomaras 2014). That relation en-sures that there is a maximum value of the turnaround radius for a structure of mass $M$, which is equal to $R_{ta,\max} = (3M/8\pi\rho_\Lambda)^{1/3}$ where $\rho_\Lambda$ is dark energy density. It means that any *non-expanding* structure of mass $M$ cannot have a radius that exceeds $R_{ta,\max}$ since this prediction is an *absolute upper limit* where its applicability neither depend on the way one determines the true turnaround radius nor from the cosmic time, nor whether the structure is considered dynamically relaxed or not. Observationally, the turnaround separates the region where the gravitational attraction of the central structure is dominant from the region where matter follows the general expansion of the Universe. Therefore, an observational test based on this requirement can establish the maximum turnaround that a massive central structure achieves at the present time. Pavlidou and Tomaras (2014) analyzing objects of different scales demonstrated that the *observed* turnaround radii are systematically smaller than those predicted by $R_{ta,\max}$. On the contrary, if an observed turnaround radius would turn out greater than $R_{ta,\max}$, this would violate the ΛCDM model. Since superclusters are expected to fit positions of maximum density contrast inside the cluster distribution, to identify their membership and mass, $R_{ta,\max}$ could be a useful analytical parameter to define their maximum extent which, however, in the future may or may not collapse. Starting from the center of each supercluster initially assumed at the position of its DPW, we estimate the mass $M_{sph}$ and its corresponding $R_{ta,\max}$ inside $n$ concentric spheres with increasing test radius $R_{sph}$ until the equality $R_{sph} = R_{ta,\max}$ will be satisfied. Subsequently, we recalculate the new center of mass of this sphere and repeat the process iteratively until the shift in the center is less than 1%. With the final center of mass, we obtain the final radius $R_{sph} = R_{ta,\max}$ and mass $M_{sph} = M_{ta,\max}$ of the non-expanding structure. Within each sphere of radius $R_{sph}$, the mass estimation has been evaluated as in Einasto et al. (2015) where the estimate is provided by the dynamical mass summation of all objects included in $R_{sph}$. Einasto et al. (2015) distinguished between two mass estimates: first, the dynamical mass summation of groups and clusters within a certain radius including also triplets and pairs evaluated by the median of each corresponding sample in order to reduce the large uncertainty (so did we). Second, they added to the first estimate the estimated mass of the intra-cluster gas (10% of the total mass) and masses of faint groups which are not detected by the limit of the T14 Catalog, but predicted by the brightest galaxies present in the region. They found that the final estimate differ from the first by a factor of 1.5, slightly smaller than the bias factor of 1.83 found by Chon et al. (2014) which adopts a scaling relation obtained from cosmological N-body simulations in order to define the bias when the dynamical masses of groups and clusters are converted in a supercluster mass. Note that that bias varies as a function of the cluster richness from 1.83 for $10^{13}$ $h^{-1}M_\odot$ limit to 3.36 for $10^{14}$ $h^{-1}M_\odot$. Since our aim does not require precise mass estimates, a simple summation of all dynamical masses of groups and clusters within each $R_{sph}$ has been applied having in mind that this method probably provides underestimated mass determinations as demonstrated by Einasto et al. (2015). Consequently, also $R_{ta,\max}$ may be underestimated providing a conservative definition of supercluster size and mass. Since $R_{ta,\max}$ represent the theoretical limit defined by the adopted mass-radius relation, clearly it cannot be used for dynamical analysis but could be useful for comparison with turnaround radii evaluated observationally. For less conservative $R_{ta,\max}$ determinations one may correct our mass estimates by a suitable bias factor given in Chon et al. (2014) as Chon et al. (2015) did in their study.

### 4.3. Results

For the CBSCL, A2142SCL, and VIR-SER SCL, the physical parameters have been quantified.
The results are listed in Table 1 where the first column lists the main parameters and units as: equatorial coord. (J2000), redshifts, comoving distances of the center of mass, $R_{ta,\max}$ the maximum turnaround radius, $M_{ta,\max}$ the mass estimates within $R_{ta,\max}$, numbers of enclosed group/clusters and notes regarding to compositions, previous identifications and references. Furthermore, in the Appendix, Table 1A shows group and cluster members classified in T14 Catalog and belonging to the CBSCL, A2142SCL, and VIR-SER SCL with a number of galaxies ≥ 10. The mutual comoving distances among supercluster centers (of masses) are: ~ 56.6 $h^{-1}Mpc$ between the CBSCL and A2142SCL, while the VIR-SER SCL is separated by ~ 97 $h^{-1}Mpc$ from the CBSCL and ~ 116 $h^{-1}Mpc$ from the A2142SCL.

**Table 1** – Main properties of the CBSCL, A2142SCL and VIR-SER SCL.

| Parameters/structure | CBSCL | A2142SCL | VIR-SER SCL |
|---|---|---|---|
| **R.A.° (J2000)** | 230° | 240° | 227° |
| **Decl.° (J2000)** | 28° | 27° | 5.8° |
| **z** | 0.075 | 0.09 | 0.0788 |
| **$d_c$ ($h^{-1}Mpc$)** | 222 | 264 | 232 |
| **$R_{ta,max}$ ($h^{-1}Mpc$)** | 19.3 | 16.4 | 20.0 |
| **$M_{(Rta,max)}$ ($h^{-1}M_\odot$)** | 5.5±2.7 x $10^{15}$ | 3.4±1.7 x $10^{15}$ | 6.1±3.0 x $10^{15}$ |
| **N° (group/cluster)** | 84 | 92 | 119 |
| **Previous identification: Author ID-number** | BS_12B, C/M_N10, E01_158, E06_761, L12_94*, L12_5372**, CM_MSCC 463 | BS_12C, E06_805, L12_001*, L12_2668**, CM_MSCC 472, E15_A2142 supercluster | BS_14, E01_154, L12_011*, L12_5390** |
| **Main Abell clusters enclosed** | A_2061, A_2065, A_2067, A_2089 | A_2142 | A_2028, A_2029, A_2033 |
| **References:** A (Abell 1958); BS (Bahcall and Soneira 1984); C/M (Cappi and Maurogordato 1992); E01 (Einasto et al. 2001); E06 (Einasto et al. 2006); L12 (Liivamagi et al. 2012; there are the two adaptive catalogs, one (*) for the main sample and (**) for the LRGs); CM (Chow-Martinez et al. 2014), E15 (Einasto et al. 2015) ||||

## 5. COMPARISON WITH OTHER STUDIES

In what follows, we compare our findings with the most relevant studies performed on the CBr in order to compare the GPM efficiency with respect to alternative methodologies (Bahcall and Soneira 1984 (BS); Postman et al. 1988 (PGH); Cappi and Maurogordato 1992 (C/M); Small et al. 1997 (S97), 1998; Kopylova and Kopylov 1998 (KK); Einasto et al. 2001 (E01); Einasto et al. 2006 (E06); Génova-Santos et al. 2010 (G-S); Liivamagi et al. 2012 (L12); Batiste and Batuski 2013 (BB); Pearson et al. 2014 (P14)); Chow-Martinez et al. 2014 (CM), Einasto et al. 2015 (E15).

### 5.1. CBSCL

It is largely accepted that the cluster composition of the Corona Borealis Supercluster includes the following Abell clusters: A2061, A2065, A2067 and A2089 (BS, C/M, PGH, KK, S97, E01, BB, P14, CM, present study). Other studies include also: A2056 (S97, P14); A2079 (BS, PGH, S97, E01, BB, P14, CM); A2092 (BS, PGH, S97, E01, KK, BB, CM) and A2124 (E01, CM). Besides, CM included also A2059, A 2073, A2106, A2122. The CBSCL center of mass lies between the two most massive clusters A2065 and A2061 (see Fig. 1). As expected, the comparison of our mass estimate of 5.5 x $10^{15}$ $h^{-1}M_\odot$ with that of P14 (0.6-12 x$10^{16}$ $h^{-1}M_\odot$), S97 (3 x $10^{16}$ $h^{-1}M_\odot$) and PGH (8 x $10^{15}$ $h^{-1}M_\odot$) is systematically underestimated confirming that supercluster mass estimated using the dynamical mass summation method turns out largely underestimated (Chon et al. 2014; Einasto et al. 2015). Also controversial is the size of the CBSCL: we have found $R_{ta,\max}$ ~ 19.3 $h^{-1}Mpc$ which is larger than ~ 12.5 $h^{-1}Mpc$ of P14, ~ 13±1.8 $h^{-1}Mpc$ of PGH and ~ 10 $h^{-1}Mpc$ of S97 (all these measurements refer to the collapsing "core" of the CBSCL). However, one should bear in mind that such discrepancies have little or null significance since $R_{ta,\max}$ is the theoretical maximum turnaround radius allowed by our adopted model so that, for definition, it would be the largest ones since an "observed" radius greater than $R_{ta,\max}$ would indicate a violation of the ΛCDM cosmological model. The maximum diameter of the CBSCL (the maximum distance between galaxies in the supercluster) was estimated by Einasto et al. (2011b) equal to 54.6 $h^{-1}Mpc$.

### 5.2. A2142SCL

A2142SCL has been studied in detail by Munari et al. (2014), Einasto et al. (2015) and Gramann et al. (2015). Its center of mass corresponds approximately to that of the Abell cluster A2142 which, in turn, is the richest cluster lying in the CBr (see Table 1 and Table 1A). Einasto et al. (2015) and Gramann et al. (2015) published detailed studies of its global density distribution dividing the supercluster into a higher-density core and lower-density outskirt regions from which stretch out a straight and extended filament (well visible in Fig. 1). On the basis of the density contrast test, they found that only the high density core region of 6-8 $h^{-1}Mpc$ radius has reached the turnaround and starts to collapse which, as expected, is lower than $R_{ta,\max}$ ~ 16.4 $h^{-1}Mpc$. The maximum diameter of the whole A2142 supercluster was found of 50.3 $h^{-1}Mpc$ (Einasto et al. 2011b, 2015). They estimated a total mass of 4.34 x $10^{15}$ $h^{-1}M_\odot$, greater than our finding of 3.4 x $10^{15}$ $h^{-1}M_\odot$ which, however, is in good accordance with their estimate of 2.9 x $10^{15}$ $h^{-1}M_\odot$ obtained from the dynamical mass summation method. That discrepancy can be overcome simply adding the mass due to intra-cluster gas and that due to undetected faint galaxy groups (Einasto et al. 2015) or, recovering the true mass estimate applying an adequate bias factor (Chon et al. 2014).

**5.3. VIR-SER SCL**

We did not find in literature specific studies on this supercluster. It is dominated by three Abell clusters: A2028, A2029 and A2033. Unexpectedly, this supercluster has comparable mass and extension of the CBSCL, but richer of galaxy groups/clusters (see Table 1). However, its total luminosity is a factor 1.5 fainter than that of the CBSCL (Einasto et al. 2011b) which provides a mass-to-light ratio of ~ 300 $h\,M_\odot/L_\odot$ significantly larger than ~ 170 $h\,M_\odot/L_\odot$ obtained for the CBSCL. Such a discrepancy confirms the suspect brought in Sec.5.1 that the mass estimated here for the CBSCL could be largely underestimated. The maximum diameter of 35.4 $h^{-1}Mpc$ given in Einasto et al. (2011b) is smaller than our estimation of 40 $h^{-1}Mpc$. The VIR-SER SCL has been previously identified as the BS_14, E01_154, L12_011. Einasto et al. (2011b) included A2040 as a supercluster member but not here since it lies at much lower redshift than the supercluster limits according to BS and E01.

**6. DOES THE CBSCL FORM A BOUND BINARY SYSTEM?**

**6.1. The hypothesis**

The large diameter of ~ 40 $h^{-1}Mpc$ measured for the CBSCL and ~ 33 $h^{-1}Mpc$ for A2142SCL and the short separation of their center of masses of ~ 56.6 $h^{-1}Mpc$ are the ingredients to speculate that they may form a bound binary system. Now, to really understand whether the CBS_A and CBS_B are expanding (outgoing) or collapsing (incoming) as well as the dynamical state of the whole structure requires a detailed analysis to verify whether the induced local velocity field is separated from large scale tidally induced flows (Courtois et al. 2012). In theory, when the dynamics is dominated by gravity, analyses of velocity fields provide precise information on the dynamical state of structures and a better basis for predicting their future evolution (Tully et al. 2014; Pomarède et al. 2015). However, at present, only for the very local region of the Universe peculiar velocity data are available. As seen in Fig. 1, 2 and 3, some observational aspects exploited by the local potential distribution seem to support the binary hypothesis. In particular, even if the filamentary structure does not fully connect the two binary members, one may recognize the similarity of that configuration with a Roche lobe potential distribution, a typical feature of a binary system. From this point of view however, it is interesting to note that the two superclusters do not overlap each other thus it is unlikely they are currently in a merging phase.

**6.2. A remark**

Our findings partially match the results of Luparello et al. (2011) which studied the future evolution of superclusters in the context of ΛCDM cosmology. They found that the CBr is dominated by two main superclusters: the SCL761 and SCL805 (ID labels of E06) that are the counterparts of the CBSCL and A2142SCL, respectively. Besides, they predicted that these structures are candidates to merge and form a single virialized system in the future. Furthermore, a similar conjecture appears in the BS Supercluster catalog. BS found that the CBSCL is an unique, large scale structure with a density enhancement factor $f$=20 which can be divided in three substructures: 12A, 12B and 12C where 12B corresponds to the CBSCL while 12C corresponds to the A2142SCL (12A region is not taken into account since its local potential well is not as deep as 12B and 12C).

**7. CONCLUSIONS**

The gravitational potential-based method (GPM) has been applied to a recent volume-limited group/cluster catalog compiled by Tempel et al. (2014) and limited to the well-studied Corona Borealis region in order to search and define the most massive structures lying in that area and unknown gravitational features unrevealed by previous analyses. Displaying the distribution of the local gravitational potentials generated by the spatial distribution of the group/cluster sample filling that region, the deepest potential wells turn out concentrated on three major clustered structures identified as the CBSCL, A2142SCL and VIR-SER SCL The CBSCL and A2142SCL are interconnected by a well-defined filamentary structure of groups and clusters forming a wide system similar to a binary supercluster. According to Luparello et al. (2011) and Bahcall and Soneira (1984) we confirm that the CBSCL is part of a giant structure including also the A2142SCL but the lack of peculiar velocity data prevents an appropriate clustering analysis to support the idea that the whole system is bound or in process of future virialization. Given the extraordinary importance of disentangle this issue, it is desirable that the peculiar velocity distribution around the two superclusters will become available to find a robust outcome with a detailed dynamical analysis.

**APPENDIX**

**TABLE 1A -** List of group/clusters with n.gal. ≥ 10 belonging to CBSCL, A2142SCL and VIR-SER SCL

| (1) | (2) | (3) | (4) | (5) | (6) | (7) |
|---|---|---|---|---|---|---|
| T14 Cat ID | n.gal. | RA° | DEC° | z | Mass $10^{12} M_\odot$ | Abell ID |
| **CBSCL** | | | | | | |
| 1004 | 16 | 231.0 | 31.11 | 0.07438 | 50.5 | A2067 |
| 1069 | 62 | 230.3 | 30.61 | 0.07862 | 609 | A2061 |
| 1568 | 25 | 233.1 | 28.05 | 0.07380 | 256 | A2089 |
| 3462 | 65 | 230.7 | 27.69 | 0.07236 | 2500 | A2065 |
| 7459 | 10 | 228 | 27.78 | 0.06988 | 114 | |
| **A2142SCL** | | | | | | |
| 1474 | 90 | 239.6 | 27.27 | 0.09028 | 1060 | A2142 |
| 1476 | 22 | 239.2 | 27.63 | 0.08937 | 219 | |
| 2413 | 14 | 239.9 | 26.56 | 0.08973 | 178 | |
| 2414 | 11 | 239.6 | 26.63 | 0.08709 | 27.1 | |
| 3380 | 13 | 240.8 | 25.41 | 0.08731 | 86.9 | |
| 5229 | 12 | 238.6 | 27.44 | 0.09172 | 52.3 | |
| 5364 | 15 | 241.2 | 24.6 | 0.08808 | 226 | |
| 7426 | 12 | 240.3 | 25.83 | 0.08837 | 66.8 | |
| 10360 | 11 | 240.8 | 26.93 | 0.09014 | 69.8 | |
| **VIR-SER SCL** | | | | | | |
| 1445 | 19 | 227.8 | 5.277 | 0.08023 | 301 | |
| 1446 | 32 | 227.9 | 6.269 | 0.08008 | 610 | |
| 1447 | 16 | 227.7 | 4.869 | 0.07985 | 188 | |
| 1448 | 49 | 227.7 | 5.766 | 0.07904 | 1670 | A2029 |
| 1587 | 10 | 226.9 | 8.655 | 0.07951 | 68.5 | |
| 2009 | 24 | 227.4 | 7.608 | 0.07828 | 285 | A2028 |
| 2254 | 10 | 231.2 | 6.943 | 0.07796 | 122 | |
| 2255 | 12 | 230.8 | 6.84 | 0.07777 | 181 | |
| 3787 | 24 | 227.4 | 8.841 | 0.0798 | 291 | |
| 4041 | 13 | 226.3 | 5.051 | 0.08121 | 51.3 | A2033 |
| 5773 | 10 | 224.5 | 8.871 | 0.08152 | 47 | |

LEGEND: column (1) ID number from T14/ M = -20 volume-limited group/cluster Catalog; (2) Number of galaxies; (3-4) J2000 equatorial coord. (5) redshift; (6) Mass in $10^{12} M_\odot$ ; (7) Abell cluster ID